\RequirePackage{fix-cm}
\documentclass[smallextended]{svjour3}       
\smartqed  
\usepackage{graphicx}
\begin{document}

\title{DFT modelling of the effect of strong magnetic field on Aniline molecule 
}


\author{H. Atci, Y. Polat, M. Huseyinoglu, \\ B. Arikan, A. Siddiki 
}


\institute{H. Atci \at
           ETH Zurich, Department of Physics, CH-8093 Zurich, Switzerland\\
           Istanbul University Physics Department 34134 Vezneciler-Fatih, Istanbul, Turkey \\
           \email{huseyinatci@gmail.com}\\         
           \and
           Y. Polat \at
              Marmara University, Biophysics Department, 34854 Maltepe, Istanbul, Turkey \\
           \and
           M. Huseyinoglu \at
              Istanbul University Physics Department 34134 Vezneciler-Fatih, Istanbul, Turkey \\
           \and   
           B. Arikan \at
              Kadir Has University, Computational Biology Department, 34083 Fatih, Istanbul, Turkey \\
           \and
           A. Siddiki \at
              Mimar Sinan Fine Arts University, Physics Department, 34380 Bomonti-Sisli, Istanbul, Turkey           
}

\date{Received: date / Accepted: date}

\maketitle

\begin{abstract}
Aniline is an organic compound with the stoichiometric expression $C_{6}H_{5}NH_{2}$; consisting of a phenyl structure attached to an amino group. It is colorless, but it slowly oxidizes and resinifies in air, giving a red-brown tint to aged samples. Until now, there are only few researches on Aniline considering low magnetic fields. In this work, we study Aniline molecule under different high magnetic fields using density functional theory methods including independent particle and interacting particle approaches. We obtain charge density distrubitions, energy dispersions, dipol moments and forces as functions of position and magnetic field. Our numerical results show that magnetic field affects electron density of the considered molecule. As a result, it is observed that there are strong fluctuations in energy dispersion.
\end{abstract}

\section{Introduction}
\label{intro}
Aniline, also known as Phenylamine or Aminobenzene, is a primary aromatic amine. A compound in which one of the Hydrogen atoms in an Ammonia molecule has been replaced by a Hydrocarbon group. As it is well known, there is a class of Hydrocarbons, which is characterized by high degree of unsaturation and unusual stability. As the most important and common member of this class is Benzene ($C_{6}H_{6}$). Aniline has an -$NH_{2}$ group attached directly to a Benzene ring. There are 6$\pi$  electrons from the Benzene ring, and an additional 2 from the $sp^{2}$ hybridized amine group. There is an interaction between the delocalised electrons in the Benzene ring and the single pair on the Nitrogen atom. The single pair overlaps with the delocalised ring electron system. The donation of the Nitrogen's single pair to the ring system increases the electron density around the ring. That makes the ring much more reactive than as it is in Benzene itself. It also reduces the availability of the single pair on the Nitrogen to take part in other reactions. In particular, this makes Aniline much more weakly basic than primary amines where the -$NH_{2}$ group isn't attached to a Benzene ring.\\ 

Aniline, $C_{6}H_{5}NH_{2}$, is colorless, oily, highly toxic liquid organic compound. Chemically as a primary aromatic Amine molecule, it is formed by replacing one Hydrogen atom of a Benzene molecule with an Amino group. Aniline boils at 184$^{o}$C; and melts at -6$^{o}$C. It is of great importance in the dye industry, being used as the starting substance in the manufacture of many dyes. For this reason many dyes have the word Aniline in their common name, such as Aniline black, Aniline red, yellow, blue, purple, orange, green, and others. Today these synthetic dyes have largely replaced the natural ones. Aniline is prepared commercially by the reduction of Nitrobenzene, originally manufactured from Benzene obtained by the distillation of coal tar, or reaction of Chlorobenzene with Ammonia in the presence of a copper catalyst. Sulfonation of Aniline yields sulfanilic acid, the parent compound of the sulfa drugs. Aniline is also important in the manufacturing of rubber-processing chemicals and antioxidants.\\

An Amino group cannot be introduced into the Benzene ring by a single chemical reaction. Instead, Nitrobenzene ($C_{6}H_{5}NO_{2}$) is formed first and is then reduced in a second reaction which yields Aniline. Studying the structures of these molecules shows that Nitrobenzene must lose Oxygen and gain Hydrogen. Aniline may also be obtained from the Nitrobenzene by reducing it, using nascent Hydrogen (i.e. Hydrogen that is actually prepared in the presence of the Nitrobenzene). This change is a reduction and the most effective laboratory reductant is a mixture of granulated tin and concentrated hydrochloric acid. Aniline is very weakly basic and the reduction medium is acidic. The nitration of Aniline is going to be faster than the nitration of Nitrobenzene, since the Aniline is a ring with $NH_{2}$ substituent and Nitrobenzene is a ring with $NO_{2}$ substituent. As described above, $NH_{2}$ is an activating group which speeds up the reaction and $NO_{2}$ is deactivating group that slows down the reaction \cite{Organik Maddeler-3}.\\ 

The influence of magnetic field on molecules is of great interest nowdays. The magnetic field influences the chemical reactions including reduction medium and can change orientation of most organic molecules. Although some literature concerning the effect of low magnetic field on Aniline exists. None of them focus on behaviour of Aniline in high magnetic fields. In this paper, we investigate the response of Aniline considering high magnetic fields. Utilizing density functional theory (DFT) methods, by the help of numerical computation, we obtain spatial charge distribution, energy dispersion, dipol moments and forces. On one hand we show that the charge distribution is strongly affected to external magnetic field at least in one spatial direction. On the other hand, we observe that the kinetic energy changes with magnetic field as a parabolic function. However, the total energy is completely nonlinear with magnetic field regarding dipol moments and forces. The numerical data presents highly nonlinear behaviour in different spatial coordinates except the z- direction. \\

The paper is organized as follows. First, we discuss our numerical methods starting from geometric optimization, which is followed by the primitive independent particle picture. Next, we summarized the basics of the DFT. In Section 3, we present our numerical results. Finally, we open our understanding and results for discussion.  

\section{Method}
\subsection{Geometric optimization}
One of the primary roles of computational chemistry is to determine the equilibrium geometry of a molecule and 
the relative energy differences between stable conformers. A potential energy surface (PES) maps out the potential
energy of a molecule as a functional of all of its geometrical parameters. The PES is multi-dimensional; for a
nonlinear molecule with N atoms, the PES will depend on 3$N$-6 geometrical coordinates \cite{Pulay1992:2856}. The local minima on the potential energy surface correspond to stable structures of the molecule. The lowest energy of these is the global minimum. Geometry optimization is the name for the procedure that attempts to locate a minimum on the potential 
energy surface in order to predict equilibrium structures of molecular systems. Mathematically, a stationary point 
is one at which the first derivative (or slope) of the potential energy with respect to each geometric parameter is 
zero:
\begin{eqnarray}\label{eqexpmuts}
\frac{\partial \textbf{E}}{\partial \textbf{q}_{1}}=\frac{\partial \textbf{E}}{\partial \textbf{q}_{2}}=...=0.
\end{eqnarray}\\

Since the configuration has also to conserve its stability under small displacements away form the staionary point, 
the stationary point has to be a local minimum of the total energy function. This means that small displacements 
away from the stationary point will always lead to an increase in energy. From a mathematical point of view a 
local minimum is characterized by the fact the the curvature along any line going through the local minimum is 
positive \cite{Pulay1992:2856}.

\begin{eqnarray}\label{eqexpmuts}
\frac{\partial^{2} \textbf{E}}{\partial \textbf{q}^{2}}>0.
\end{eqnarray}

The geometric optimization of the structures are performed using density funtional theory (DFT) Becke’s 
three-parameter  \cite{Becke1993:5648} hybrid function with the non-local correlation of Lee-Yang-Parr (B3LYP) method \cite{Lee1988:785}. All of calculations are using 6-31+G(d) basis set with help of Gaussian03 package program  \cite{Gaussian03}.

\subsection{Independent particles}
In this section we will go through a very important concept: the independent particle approximation. This 
approximation is at the heart of many methods such as Hartree-Fock theory and DFT which are very fundamental methods 
to solve the electronic Schrodinger equation: 
\begin{eqnarray}\label{eqexpmuts}
H\Psi_{i}=E\Psi_{i}
\end{eqnarray}
The first assumption within the independent particle approximation is each particle is in a different orbital, so that we can write the wavefunction in a product form \cite{Kohn1965:A1133}:
\begin{eqnarray}\label{eqexpmuts}
\phi(r_{1},r_{2},...,r_{i})=\eta_{1}(r_{1})\eta_{2}(r_{2})...\eta_{N}(r_{N})  
\end{eqnarray}
where $\eta_{i}$ is the number of $N$ orbitals. This equation is a useful independent particle approximation for the wavefunction and means that electron 1, whose position is given by $r_{1}$, is in the orbital $\eta_{1}$, electron 2 whose 
position is given by $r_{2}$, is in the orbital $\eta_{2}$ and so on. 

\subsection{Density functional theory}
The DFT is presently the most successfull approach to compute the electronic structure of matter. Its applicability ranges from atoms, molecules, solids to nuclei more over to classical and quantum fluids. In its original formulation, DFT provides the ground state properties of a system, and the electron density plays a key role. As an example, DFT predicts great variety of molecular properties: molecular structures \cite{Ropo2014:241401(R)}, atomization energies \cite{Haas2010:125136}, ionization energies \cite{Hellgren2013:013414}, magnetic properties \cite{Atci2016:00113}, \cite{Atci2013:155604}, dipol moment \cite{Rashin1994:182}, force \cite{Reine2010:044102}, etc. The original DFT has been generalized to deal with many different situations: spin polarized systems \cite{Atci2013:155604}, multicomponent systems \cite{Culpitt2016:044106}, time dependent phenomena and excited states \cite{Salman2013:155}, molecular dynamics \cite{Aichinger2010:016703}, etc. \\

The ultimate goal of most approaches in solid state physics and quantum chemistry is the solution of the time-independent, non relativistic Scrodinger equation,
\begin{eqnarray}\label{eqexpmuts}
\widehat{H}\Psi_{i}(\vec{x}_{1},\vec{x}_{2},...,\vec{x}_{M},\vec{R}_{1},\vec{R}_{2},...,\vec{R}_{M})=E\Psi_{i}(\vec{x}_{1},\vec{x}_{2},...,\vec{x}_{M},\vec{R}_{1},\vec{R}_{2},...,\vec{R}_{M})
\end{eqnarray} 
$\widehat{H}$ is the Hamiltonian for a system consisting of $M$ nuclei and $N$ electrons. 
\begin{equation}\label{eqexpmuts}
\widehat{H}=-\frac{1}{2}\sum^N_{i=1}\bigtriangledown^2_{i}-\frac{1}{2}\sum^M_{i=1}\frac{1}{M_{A}}\bigtriangledown^2_{A}-\sum^N_{i=1}\sum^M_{i=1}\frac{Z_{A}}{r_{iA}}+\sum^N_{i=1}\sum^N_{j>i}\frac{1}{r_{ij}}+\sum^M_{A=1}\sum^M_{B>A}\frac{Z_{A}Z_{B}}{R_{AB}}
\end{equation}
Here, $A$ and $B$ run over the $M$ nuclei while $i$ and $j$ denote the $N$ electrons in the system. The first two terms describe the kinetic energy of the electrons and nuclei. The other three terms represent the attractive electrostatic interaction between the nuclei and the electrons and repulsive potential due to the electron-electron and nucleus-nucleus interactions.\\
 
The electron density is the central quantity in DFT. It is defined as the integral over the spin coordinates of all electrons and over all but one of the spatial variables (\vec{x\equiv\vec{r} },s )
\begin{eqnarray}\label{eqexpmuts}
\rho(\vec{r})= N\int ...\int \arrowvert \Psi(\vec{x}_{1},\vec{x}_{2},...,\vec{x}_{N}) \arrowvert^{2}d\vec{x}_{1}\vec{x}_{2}...\vec{x}_{N}
\end{eqnarray}
$\rho(\vec{r})$ is a non-negative function of only the three spatial variables which vanishes at infinity and integrates to the total number of electrons:
\begin{eqnarray}\label{eqexpmuts}
\rho(\vec{r\rightarrow \infty }),\\
\int \rho(\vec{r})d\vec{r}=N. 
\end{eqnarray}
$\rho(\vec{r})$ is an observable and can be measured experimentally, e.g. by X-ray diffraction.\\

The conventional approaches use the wave function $\Psi$ as the central quantity, since $\Psi$ contains the full information of a system. However, $\Psi$ is a very complicated quantity that cannot be probed experimentally and that depends on $4N$ variables, $N$ being the number of the electrons. The energy is given completely in terms of the electron density. The resulting equations are the Kohn-Sham equations \cite{Kohn1965:A1133},\cite{Barth1972:1629}:
\begin{eqnarray}\label{eqexpmuts}
\biggr(-\frac{1}{2}\nabla ^{2}+\biggr[\int \frac{\rho (\vec{r}_{2} )}{r_{12}}+V_{XC}(\vec{r}_{1})-\sum^M_{A}\frac{Z_{A}}{r_{1A}}\biggr]\biggr)\Psi _{i}=\epsilon _{i}\Psi _{i}\\
V_{KS}(\vec{r}_{1})=\int \frac{\rho (\vec{r}_{2} )}{r_{12}}+V_{XC}(\vec{r}_{1})-\sum^M_{A}\frac{Z_{A}}{r_{1A}}.
\end{eqnarray}
Once we know the various contributions in Eqs. (10-11) we have a grip on the potential $V_{KS}$ which we need to insert into the one-particle equations, which in turn determine the orbitals and hence the ground state density and the ground state energy. Notice that $V_{KS}$ depends on the density, and therefore the Kohn-Sham equations have to be solved iteratively. The exchange-correlation potential $V_{XC}$ is defined as the functional derivative of $E_{XC}$ with respect to $\rho$, i.e. $V_{XC}=\delta E_{XC}/\delta \rho$. It is very important to realize that if the exact forms of $V_{XC}$ and $E_{XC}$ were known, the Kohn-Sham strategy would lead to the exact energy. To calculate exchange-correlation energy $E_{XC}$ , we use the local spin density approximation with a parametrization provided by Attaccalite et al \cite{Attaccalite2002:256601}. In the spin DFT calculations, we utilise the OCTOPUS code package \cite{Marques2003:60} (published under the General Public License) built on the real space grid discretization method which allows realistic modeling of two dimensional systems. Related technical details can be found in Reference [19]. To solve Schrodinger equation the conjugated gradient algorithm is used. 

\section{Results}

\begin{figure}
\includegraphics[width=1.0\textwidth]{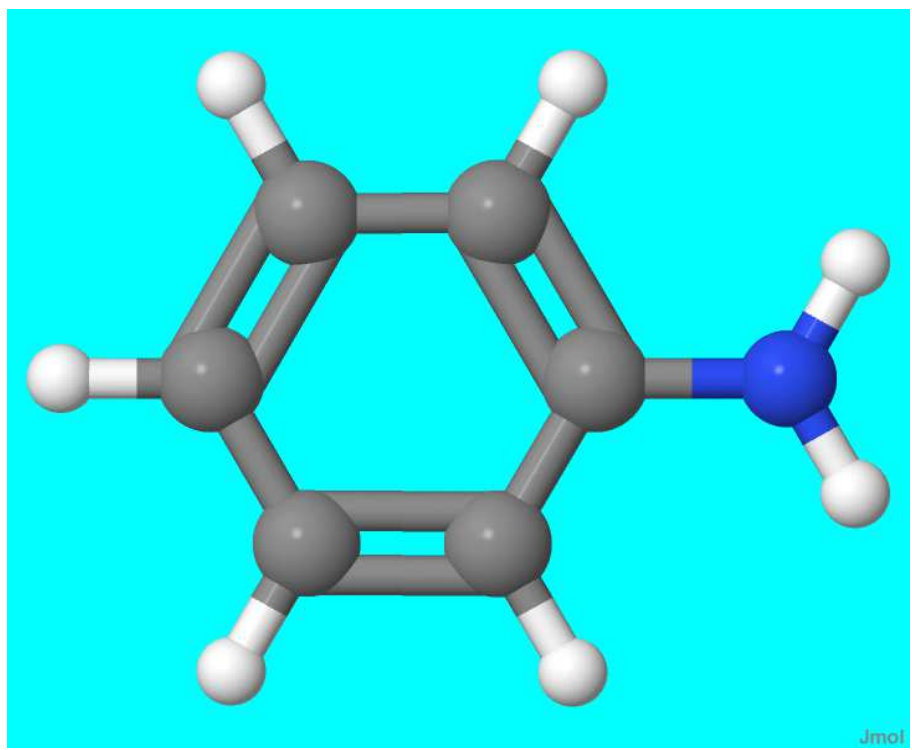}
\caption{Schematic view of an Aniline molecule, taken from Jmol program \cite{Jmol}. The molecules are organized as follows: white: Hydrogen, dark (black): Carbon, darker (blue): Nitrogen. There are single and double bonds in Benzene ring.}
\label{fig:1}       
\end{figure}

In an attempt to explain the biological effects of static magnetic fields, it is useful to classify them as weak (\textless 1mT), moderate (1mT to 1T), strong (1-5T) and ultrasong (\textgreater5T) \cite{Dini2005:195}. In our calculations we consider moderate, strong and ultrastrong magnetic fields using independent particles and DFT methods.\\ 

The system we study is shown in Fig. 1, Aniline molecule consists of five Carbon atoms, seven Hydrogen atoms and one Nitrogen atom. In Fig. 1, we show schematic view of Aniline using Jmol program \cite{Jmol}.\\ 

The intramolecular magnetic field around an atom in a molecule changes the electron density, thus giving access to details of the electronic structure of a molecule. The main reason is the influencing of the magnetic component of spin angular momentum of electrons. It is observed it affects the entire electron wave functions under the influence of a magnetic field given from the outside of this component and is shown how this change is at different magnetic field intensity in the figures. Fig. 2 and Fig. 3 show that comparison of calculated densities considering different magnetic fields using independent particles method in x-axis and y-axis, respectively.\\

\begin{figure*}
\includegraphics[width=1.00\textwidth]{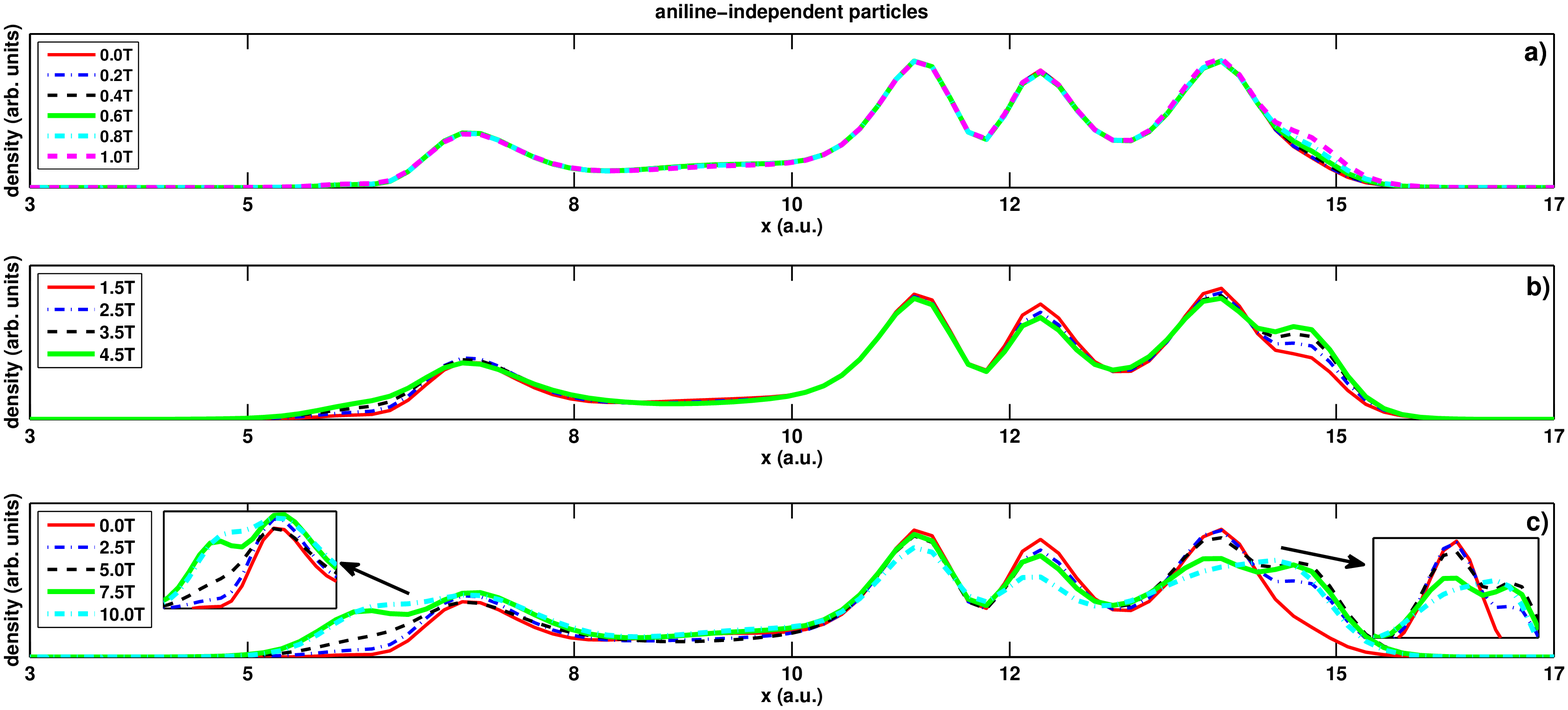}
\caption{Comparison of calculated densities considering different high magnetic fields using independent particles method in the x-axis a) at low b) strong c) ultrastrong magnetic fields. Inset shows the differences arising from applying different magnetic fields.}
\label{fig:2}       
\end{figure*}

\begin{figure*}
\includegraphics[width=1.00\textwidth]{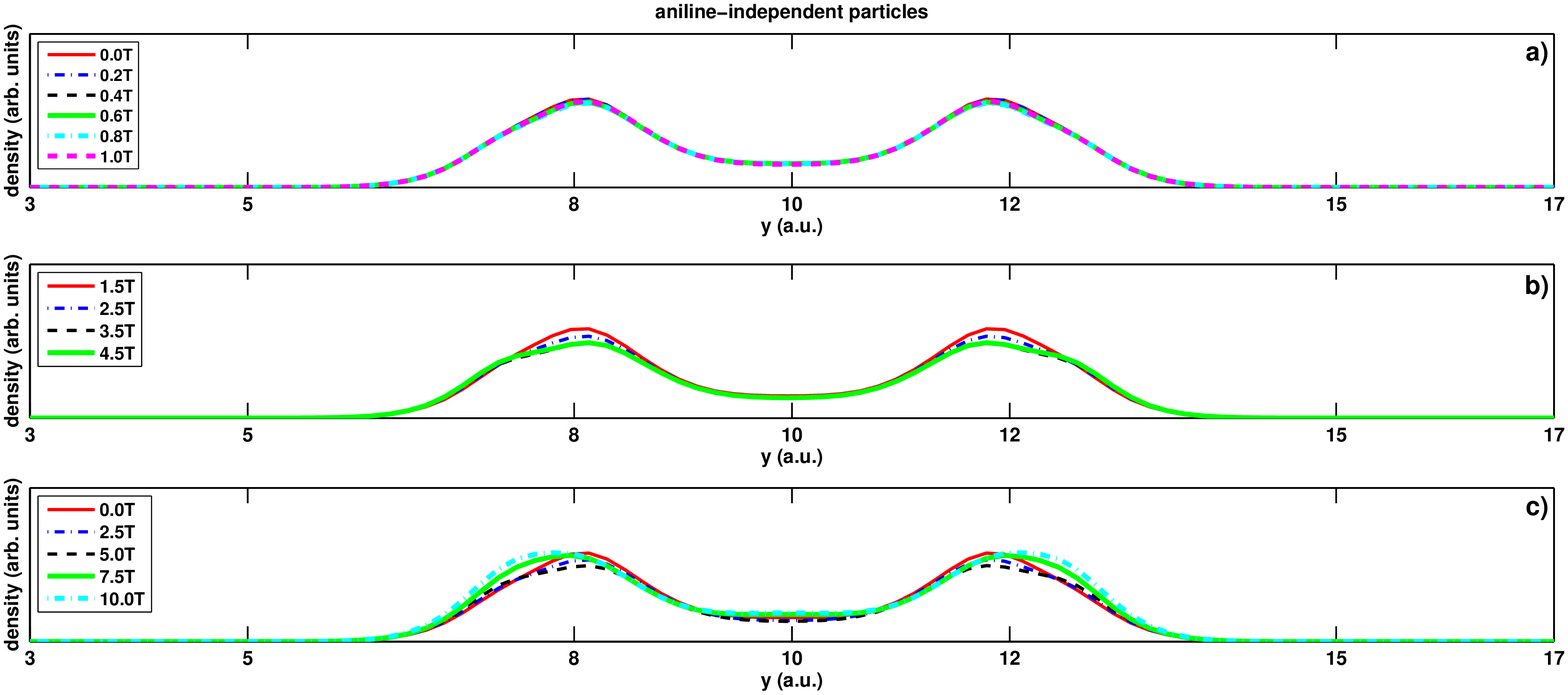}
\caption{Comparison of calculated densities considering different high magnetic fields using independent particles method in the y-axis a) at low b) strong c) ultrastrong magnetic fields.}
\label{fig:3}      
\end{figure*}

In Fig. 4 and Fig. 5 we present the calculated charge densities under different magnetic field using DFT method in x-axis and y-axis, respectively. We see a weak static magnetic field has no effect on Aniline molecule but it has also been shown that Aniline is influenced by extremely strong magnetic fields (up to maximum 10T) and there are more differences between weak and strong magnetic fields. Due to the change of density is directly related to electron energies, we observe a change in energy and this is a common situation.\\

\begin{figure*}
\includegraphics[width=1.00\textwidth]{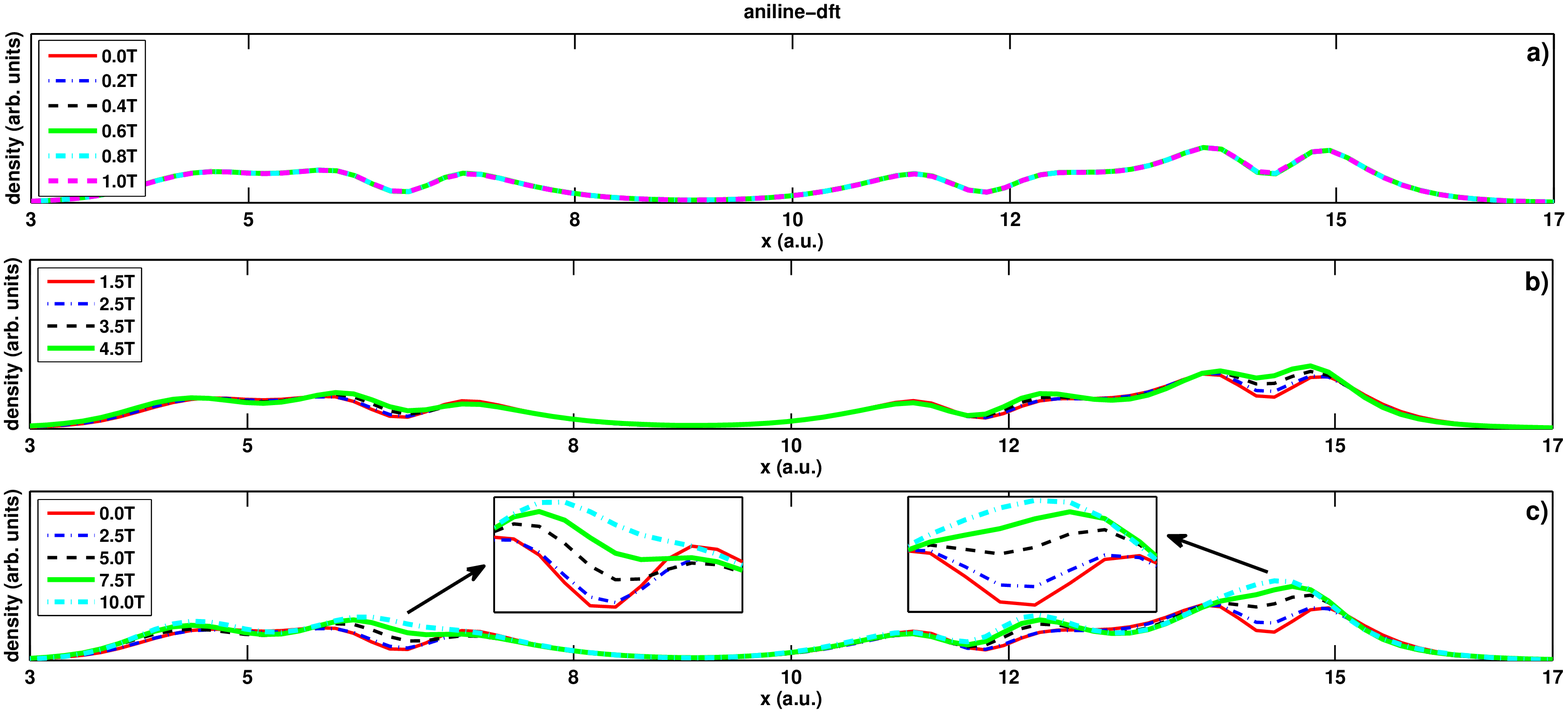}
\caption{Comparison of calculated densities considering different high magnetic fields using DFT method the in the x-axis a) at low b) strong c) ultrastrong magnetic fields. Inset shows the differences arising from applying different magnetic fields.}
\label{fig:4}     
\end{figure*}

\begin{figure*}
\includegraphics[width=1.00\textwidth]{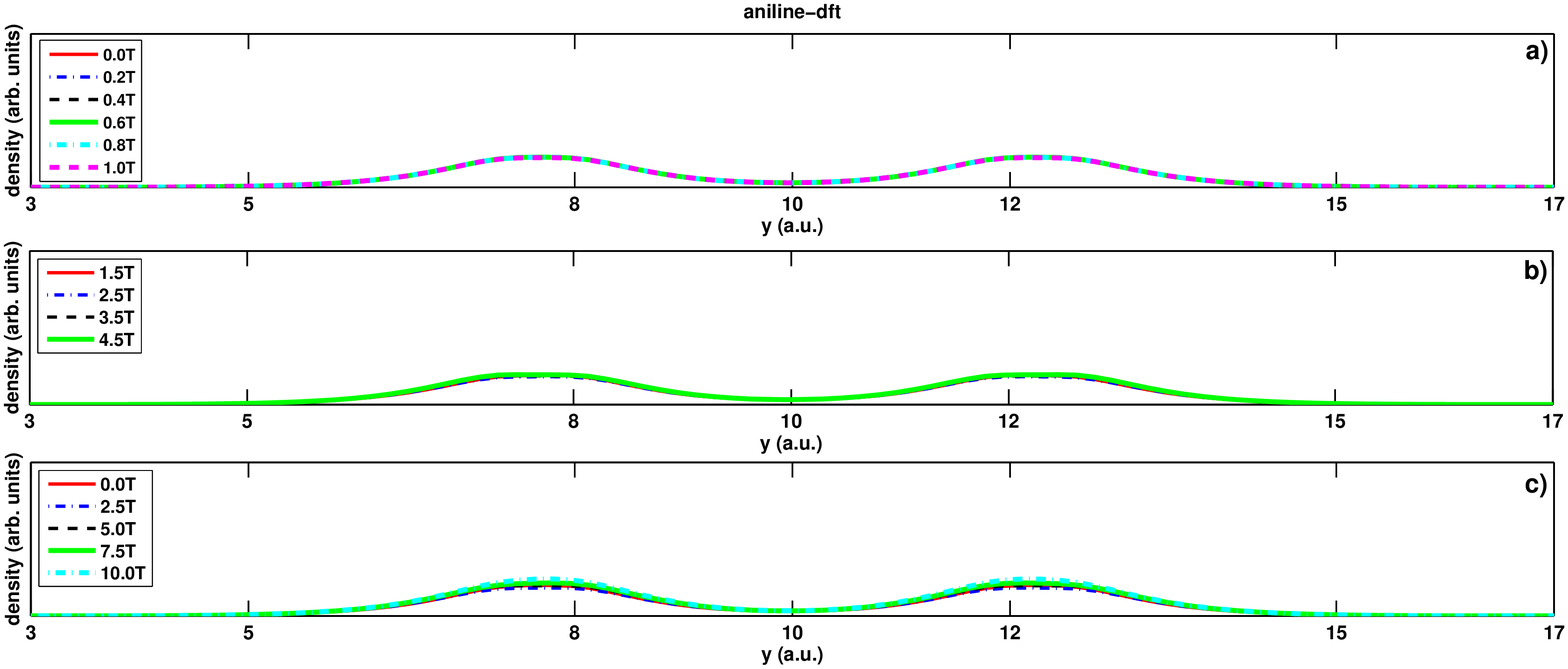}
\caption{Comparison of calculated densities considering different high magnetic fields using DFT method in the y-axis a) at low b) strong c) ultrastrong magnetic fields.}
\label{fig:5}    
\end{figure*}

In Fig. 6, we comparise normalized energies at 0T versus magnetic field and we observe that there has been a fluctuation in energies values between 2T and 5T. Energy parabolic curve (Hartree, exchange, and correlation), the rise around 2T become descent by a sudden release, then return to its original orientation in the vicinity of 5T. This is why energy release brings the prediction that the charge of the electron density distribution is directly proportional to the size of a threshold magnetic components. Our external magnetic field strength is the reason that it provides an undulating change in energy when passing this threshold, and this change allows normal rotation of the magnetic field strength increases. This is because increasing of the energy is back to normal with the rise above the threshold value of the external magnetic field. The total energy situation is exactly the opposite, and this time it switches to decrease. Inset of Fig. 6 shows normalized kinetic energy versus magnetic field increases.\\ 

\begin{figure*}
\includegraphics[width=1.05\textwidth]{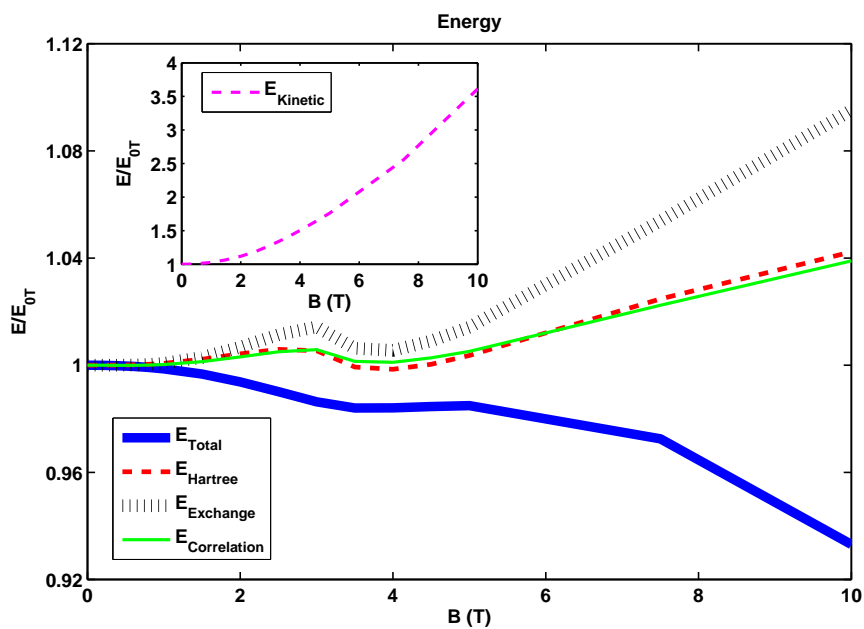}
\caption{Comparision of normalized energies at 0 Tesla ($E_{0T}$) versus magnetic field. Inset shows normalized kinetic energy versus magnetic field.}
\label{fig:6}     
\end{figure*}

In Fig. 7, we show the calculated dipole moments in x-, y-, z- axis using DFT method. The applied magnetic field induces circulations in the electron cloud surrounding the nucleus such that a magnetic moment $\mu$, opposed to magnetic field, is produced. Nuclei in a region of high electron density are more shielded from the applied field than those in regions of lower electron density. If inductive effects are present in a molecule, reduction of the electron density and dishielding is expected. Chemical bonds are regions of high electron density and therefore can create local magnetic fields. Electrons within bonds are not usually able to circulate freely and so the chemical shift will depend on the orientation of the nucleus with respect to the bond. \\

\begin{figure*}
\includegraphics[width=1.05\textwidth]{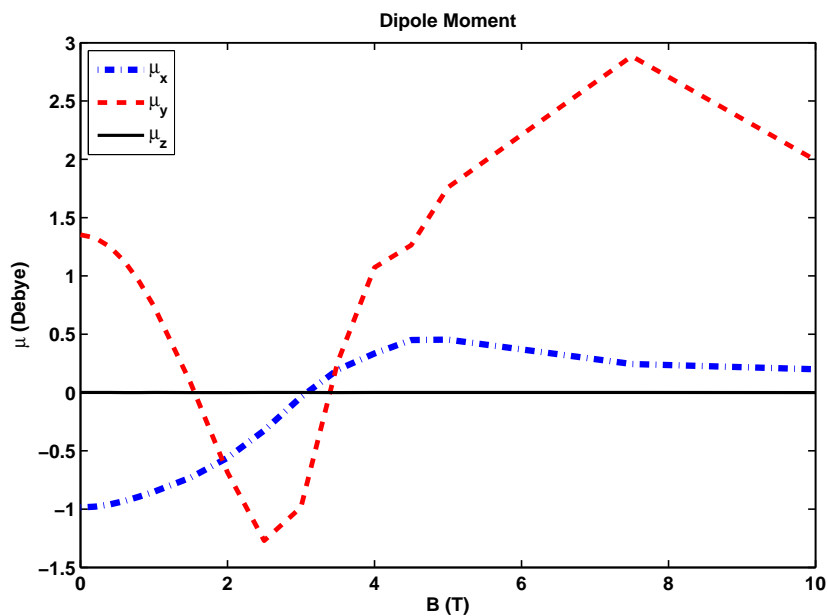}
\caption{Comparison of dipole moments in the x, y, z variables. The variables are organized as follows: x variable is smoothly increasing and decreasing, respectively. y variable is decreasing, increasing and decreasing, respectively and there are two picks. z variable is constant and continuous linear.}
\label{fig:7}  
\end{figure*}

\begin{figure*}
\includegraphics[width=1.05\textwidth]{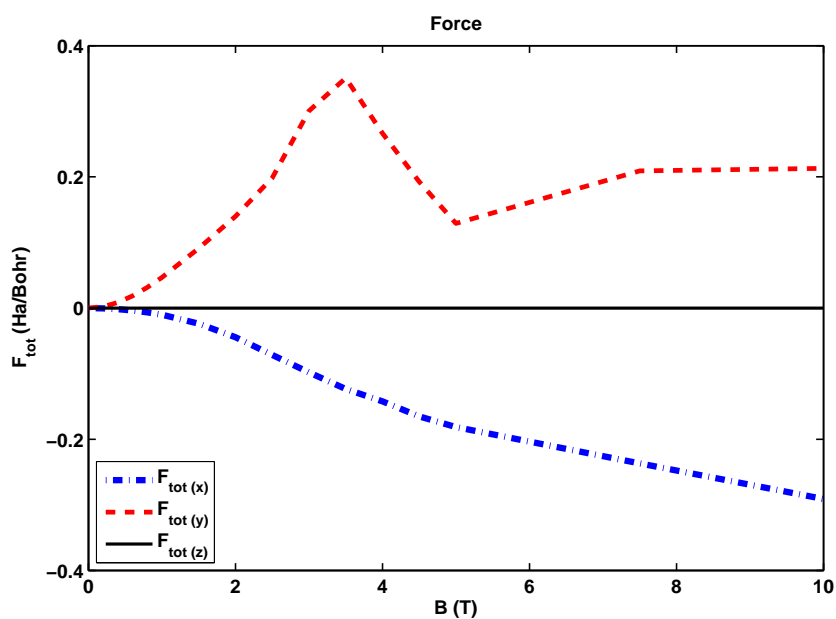}
\caption{Comparison of forces in the x, y, z variables. The variables are organized as follows: x variable is continuosly decreasing. y variable is increasing, decreasing and increasing, respectively and there is a pick. z variable is constant and continuous linear.}
\label{fig:8}    
\end{figure*}

In Fig.8, we calculate forces in x-, y-, z- axis using DFT method. Coulombic forces between the electrons and the nuclei determine the radical’s geometry and electronic structure. These, in turn, produce contributions to the radical spin Hamiltonian characterized by the g-tensor and hyperfine tensors.\\

\section{Summary}
The present information of the affects of magnetic field on Aniline is still rather primitive. Although an increasing interest and research are conducted which do result in couple of number of reports, the understanding of the field effects on the biological systems needs to be improved. The effects that have been attributed to strong and ultrastrong magnetic fields are related to their tendency to alter the preferred orientation of a variety of diamagnetic anisotropic organic molecules. \\

In this work, we think that we contributed using DFT methods considering various magnetic fields starting from simple geometric optimization. For sure it is a long way to understand the behaviour of biological molecules under high magnetic fields, however, our approach paves the path at a reasonable and consistent manner. Experimental investigations of biological molecules under strong and ultrastrong magnetic fields would lead to a better understanding which might be useful for the molecular modelling community. 

\begin{acknowledgements}
This work was supported by the Scientific and Technical Research Council of Turkey (TUBITAK) under grant no: 2214/A (H.A.), TUBITAK under grant no: 2209/A (Y.P and M.H.), TUBITAK under grant no: TBAG-112T264 (H.A. and A.S.) and TUBITAK under grant no: 211T148 (A.S.).
\end{acknowledgements}

\end{document}